\documentclass{PoS}

\usepackage{amssymb,amsmath,wrapfig}
  
\usepackage{dsfont}
\usepackage{bbold}
\usepackage{slashed}
\usepackage{url}

\title{Loop formulation of the supersymmetric nonlinear O$(N)$ sigma model}

\ShortTitle{Loop formulation of the SUSY nonlinear O$(N)$ sigma model}

\author{\speaker{Kyle Steinhauer} and Urs Wenger \\
        Albert Einstein Center for Fundamental Physics\\
         Institute for Theoretical Physics\\
        University of Bern\\
        Sidlerstrasse 5\\
        CH--3012 Bern\\
        Switzerland\\
        E-mail:  \email{steinhauer@itp.unibe.ch}, \email{wenger@itp.unibe.ch}}

      \abstract{We derive the fermion loop formulation for the
        supersymmetric nonlinear O($N$) sigma model by performing a
        hopping expansion using Wilson fermions. In this formulation
        the fermionic contribution to the partition function becomes a
        sum over all possible closed non-oriented fermion loop
        configurations. The interaction between the bosonic and
        fermionic degrees of freedom is encoded in the constraints
        arising from the supersymmetry and induces flavour changing
        fermion loops. For $N \ge 3$ this leads to fermion loops which
        are no longer self-avoiding and hence to a potential sign
        problem. Since we use Wilson fermions the bare mass needs to
        be tuned to the chiral point. For $N=2$ we determine the
        critical point and present boson and fermion masses in the
        critical regime.  }

\FullConference{31st International Symposium on Lattice Field Theory - LATTICE 2013\\
		July 29 - August 3, 2013\\
		Mainz, Germany}

\newcommand{\be}{\begin{equation}}
\newcommand{\ee}{\end{equation}}

\newcommand{\beann}{\begin{eqnarray*}} 
\newcommand{\eeann}{\end{eqnarray*}}

\begin{document}

\section{Motivation}
Despite the fact that nature has so far not revealed any trace of
supersymmetry in its elementary particle spectrum, supersymmetric
quantum field theories remain to be fascinating objects to study {\it
  per se}. It is for example most interesting to examine various
discretisation schemes for supersymmetric field theories regularised
on the lattice, e.g.~using twisted supersymmetry or or\-bi\-folding
techniques, in order to understand how the supersymmetry is realised
in the continuum limit.  Moreover, recent developments in simulating
supersymmetric field theories in low dimensions efficiently and
without critical slowing down \cite{Wenger:2008tq,Wenger:2009mi}
brings the non-perturbative study of these theories to a new,
unprecedented level of accuracy, allowing for example precise
investigations of spontaneous supersymmetry breaking phase transitions
\cite{Baumgartner:2011jw,Wenger:2012jf}.  In these proceedings we
report on our ongoing study to apply such a programme to the
supersymmetric nonlinear O$(N)$ sigma model regularised on the
lattice. This model has already been investigated numerically using a
variety of different discretisations
\cite{Catterall:2003uf,Catterall:2006sj,Flore:2012xj}. Here we
concentrate on reformulating the model in terms of fermion loops in
order to make use of the efficient simulation algorithms
and related methods to control the fermion
sign problem accompanying any spontaneous supersymmetry breaking
\cite{Baumgartner:2011cm}.

\section{Continuum model}
The Lagrangian density of the supersymmetric nonlinear O$(N)$ sigma
model in two-dimensional Euclidean spacetime, originally derived in
\cite{Witten:1977xn,DiVecchia:1977bs}, can be written as
\begin{equation}
\label{ONlagrange}
 \mathcal{L}=\frac{1}{2g^2}\left(\partial_\mu \phi \partial^\mu \phi + i \overline{\psi}\slashed{\partial}\psi + \frac{1}{4}(\overline{\psi}\psi)^2\right)
\end{equation}
where $\phi$ is a $N$-tuple of real scalar fields, $\psi$ a
$N$-tuple of real Majorana fields with
$\overline{\psi}=\psi^T\mathcal{C}$ and $\mathcal{C}$ the charge
conjugation matrix.  For the action to be O$(N)$-invariant and
supersymmetric the fields must fulfill the constraints
\begin{equation}
\label{constraints}
 \phi^2 = 1 \quad \quad \text{and} \quad \quad \phi \psi=0 \, .
\end{equation}
The model described in eq.(\ref{ONlagrange}) and the constraints in
eq.(\ref{constraints}) are both invariant under the $\mathcal{N}=1$
supersymmetry transformations
\begin{equation*}
  \delta \phi = i \overline{\epsilon}\psi \quad \text{and} \quad \delta \psi = (\slashed{\partial}+\frac{i}{2}\overline{\psi}{\psi})\phi \epsilon 
\end{equation*}
where $\epsilon$ is a constant Majorana spinor.  There is an
additional $\mathbb{Z}_2$ chiral symmetry realised by $\psi \to i
\gamma_5 \psi$ with $\gamma_5=i\gamma_0\gamma_1$.  As shown in
\cite{Polyakov:1975rr} the one-loop $\beta$-function coincides with
the one calculated for the model without SUSY, so it is asymptotically
free for $N\geq3$. As pointed out in \cite{Zumino:1979et} there exists
an $\mathcal{N}=2$ extension for supersymmetric nonlinear sigma models
which have a K\"ahler target manifold. This is the case for $N=3$ and
hence an additional SUSY can be worked out
\cite{Flore:2012xj}.  The constraints in eq.(\ref{constraints}) are
implemented in the partition function by inserting adequate Dirac
delta functions in the integration measure,
\begin{align*}
\label{ON:partitionfunction:continuum}
Z=&\int \mathcal{D} \phi \, \delta(\phi^2-1) \int \mathcal{D}\psi\, \delta(\phi\psi)e^{-S(\phi,\psi)} \\
=&\int \mathcal{D} \phi  \, \delta(\phi^2-1)e^{-S_B(\phi)} \int \mathcal{D}\psi \left( \sum_{\substack{\\ i\leq j}}^{N} \phi_{i}\phi_{j}\overline{\psi}_{i}\psi_{j} \right)e^{-S_F(\psi)} \, ,
\end{align*}
where the fermionic measure is rewritten in the second line using the
Grassmann integration rules for Dirac delta functions. In this form it
is explicit that the fermionic constraint $\phi \psi=0$ induces
flavour-diagonal ($i=j$) and flavour-changing ($i\neq j$) interactions
between the bosonic and fermionic degrees of freedom.

\section{Loop formulation and sign problem}
\label{sec:loop_formulation}
When regularising the model on a discrete space-time lattice we follow
the strategy in \cite{Flore:2012xj} and use the Wilson derivative for
both the fermionic and the bosonic action. This strategy is well
supported by various theoretical and numerical arguments
\cite{Baumgartner:2011jw,Wenger:2012jf,Baumgartner:2011cm,Golterman:1988ta,
  Catterall:2003ae,Bergner:2007pu,Kastner:2008zc,Wozar:2011gu,
  Baumgartner:2012np}.  We can then perform a
hopping expansion to all orders for the fermionic and the bosonic
variables in order to obtain an exact reformulation of the partition
function in terms of non-oriented, closed fermion loops and
constrained bosonic bond configurations (boson loops)
\cite{Baumgartner:2011cm}. These loops can in principle be simulated
effectively by enlarging the configuration space by an open fermionic
string \cite{Wenger:2008tq,Wenger:2009mi} and bosonic worms
\cite{Prokof'ev:2001zz}, respectively.  The fermion loop formulation in particular
solves the fermion sign problem since it naturally decomposes the
contributions to the partition function into bosonic and fermionic
ones, each with a definite sign depending on the employed boundary
conditions \cite{Baumgartner:2011cm}. The control of these signs is
particularly important in the context of spontaneous supersymmetry
breaking when the partition function for periodic boundary conditions,
i.e.~the Witten index, vanishes due to the exact cancellation of the
bosonic and fermionic contributions.
\begin{figure}[t]
\begin{center}
\includegraphics[scale=1.0]{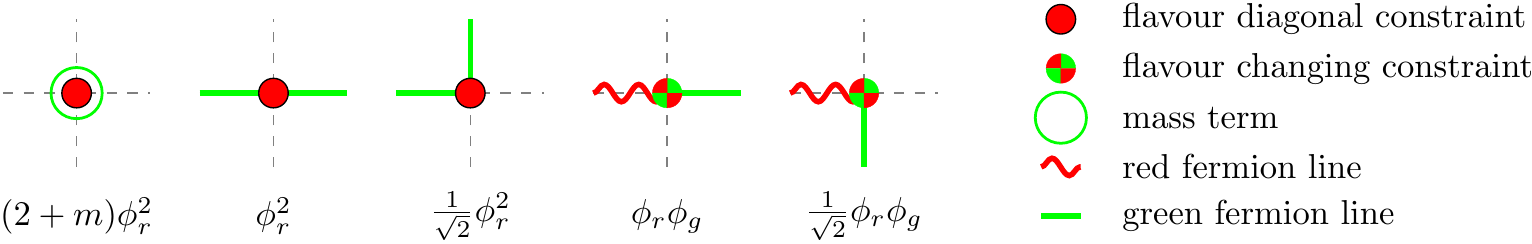}%
\caption{All possible vertices for $N=2$ up to rotations and exchange
  of the two flavours denoted as $r$ (red) and $g$ (green) with the
  corresponding weights.}
\label{fig_allvortices}
\end{center}
\vspace{-0.5cm}
\end{figure}

Unfortunately, the model as discretised above nevertheless suffers
from a sign problem. Firstly, the fermionic constraint $\phi\psi=0$
requires on each lattice site exactly one term of the form
$\phi_{i}\phi_{j}\overline{\psi}_{i}\psi_{j}$ inducing the
interactions between fermions and bosons. Since the
flavour-nondiagonal constraint with $i \ne j$ changes the flavour
content within a fermion loop, a loop can be self-intersecting for $N
\ge 3$ and hence looses the crucial property of having a definite sign
depending only on its winding topology. Secondly, the Wilson term
introduced in the bosonic sector generates a next-to-nearest neighbour
diagonal hopping term with the wrong sign leading to an overall sign
which fluctuates under local changes of the bosonic bond
configuration. In order to avoid these complications we restrict
ourselves in the following to $N=2$ for which the fermion loops remain
self-avoiding and in addition treat the bosonic degrees of freedom in
the standard way, i.e., without employing a hopping expansion. The
partition function for a system with $V$ lattice sites can then be
written as
\[
Z = \left(\frac{1}{g^2}\right)^V \int {\cal D} \, \phi \,
\delta(\phi^2-1) \,  e^{-S_B(\phi)} \sum_{\{l\}}
w(l,\phi)
\]
where the weight $w(l,\phi)$ for a given loop configuration $l$ is
determined solely by the geometry of the loops and, through the
constraints in eq.(\ref{constraints}), depends on the bosonic field
$\phi$.  As an interesting side remark we note that the coupling $g$
has no influence on the fermion loops except indirectly through the
bosonic fields.  In Figure \ref{fig_allvortices} we show all allowed
vertices up to rotations and exchange of the two flavours denoted as
$r$ (red) and $g$ (green). The additional factors are $1/\sqrt{2}$
from the Dirac algebra structure for each corner in the loop
\cite{Wolff:2007ip} and $(2+m)$ from the fermionic monomer term (with
$m$ being the bare fermion mass).  Note that the weights corresponding
to the flavour changing constraints are still not guaranteed to be
positive definite since the bosonic field components appear linearly
in those. For certain values of the parameters $m$ and $g$ this leads
to a sign fluctuating under local changes of the bosonic field and
this is why we consider the phase quenched model with $w(l,\phi)
\rightarrow |w(l,\phi)|$ in the following. The correct model can then
be recovered by reweighting with the sign $\sigma(w)$ and this seems
to work well in some of the interesting parameter regimes.

\section{Results for $N=2$}
The supersymmetric nonlinear O$(N)$ sigma model is defined in the
chiral limit where both the fermions and bosons are massless. However,
since the Wilson term explicitly breaks chiral symmetry even at zero
bare fermion mass, the chiral limit of the regularised theory is not
defined simply by the vanishing of that mass. Instead, it needs to
be tuned to the critical value $m_c$ where the correlation length of
the fermion diverges and the fermion develops a zero mode yielding
$Z_{pp}(m_c)=0$ for the partition function with periodic boundary
\begin{figure}[t]
\begin{center}
\includegraphics{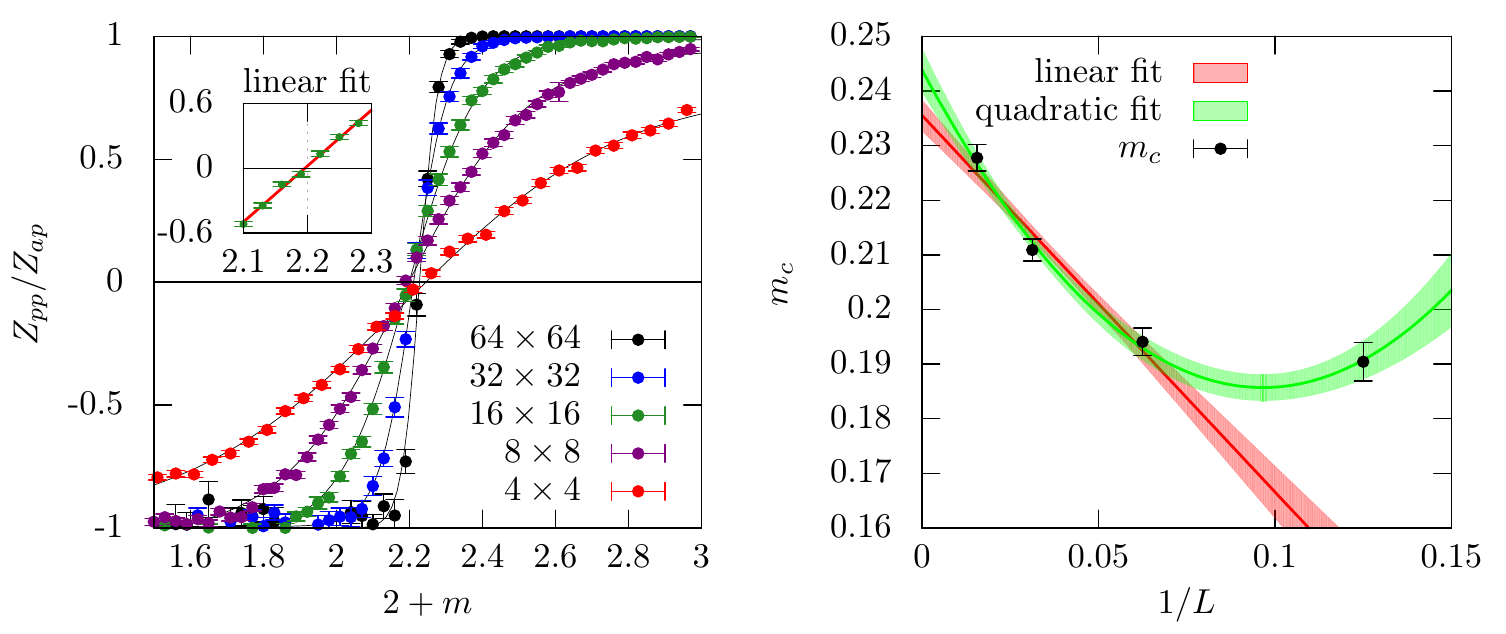}%
\caption{{\it Left plot}: $Z_{pp}/Z_{ap}$ as a function of the bare
  mass $m$ at $g=0.5$ for various lattice extents. The small inset
  shows a linear fit used to obtain the critical mass defined by
  $Z_{pp}(m_c)=0$.  {\it Right plot}: Critical masses $m_c(L)$ at $g=0.5$ for
  various lattice extents together with an extrapolation to the
  thermodynamic limit.}
\label{fig_Zratios}
\end{center}
\vspace{-0.5cm}
\end{figure} 
 conditions in both directions \cite{Wolff:2007ip,Bar:2009yq}.

 The behaviour of $Z_{pp}/Z_{ap}$ when varying the bare mass is
 illustrated in the left panel of fig.~\ref{fig_Zratios} for various
 lattice extents $L$ at $g=0.5$. The inset illustrates for the $16
 \times 16$ lattice how the values of $m_c(L)$ are extracted by
 performing linear fits to $Z_{pp}/Z_{ap}$ in the region close to
 zero. It is evident that this definition of the critical mass allows
 rather accurate determinations of $m_c$ and hence leads to precise
 extra\-polations to the thermodynamic limit. In the right panel of
 fig.~\ref{fig_Zratios} we show such an extrapolation at fixed
 coupling $g=0.5$. Eventually, this procedure can be repeated for a
 range of couplings in order to obtain $m_c(g)$ in the thermodynamic
 limit. We note however that the behaviour of the system changes
 significantly for $g \gtrsim 0.8$ where we expect the occurrence of a
 Kosterlitz-Thouless phase transition.

\begin{figure}[t]
\begin{center}
\includegraphics{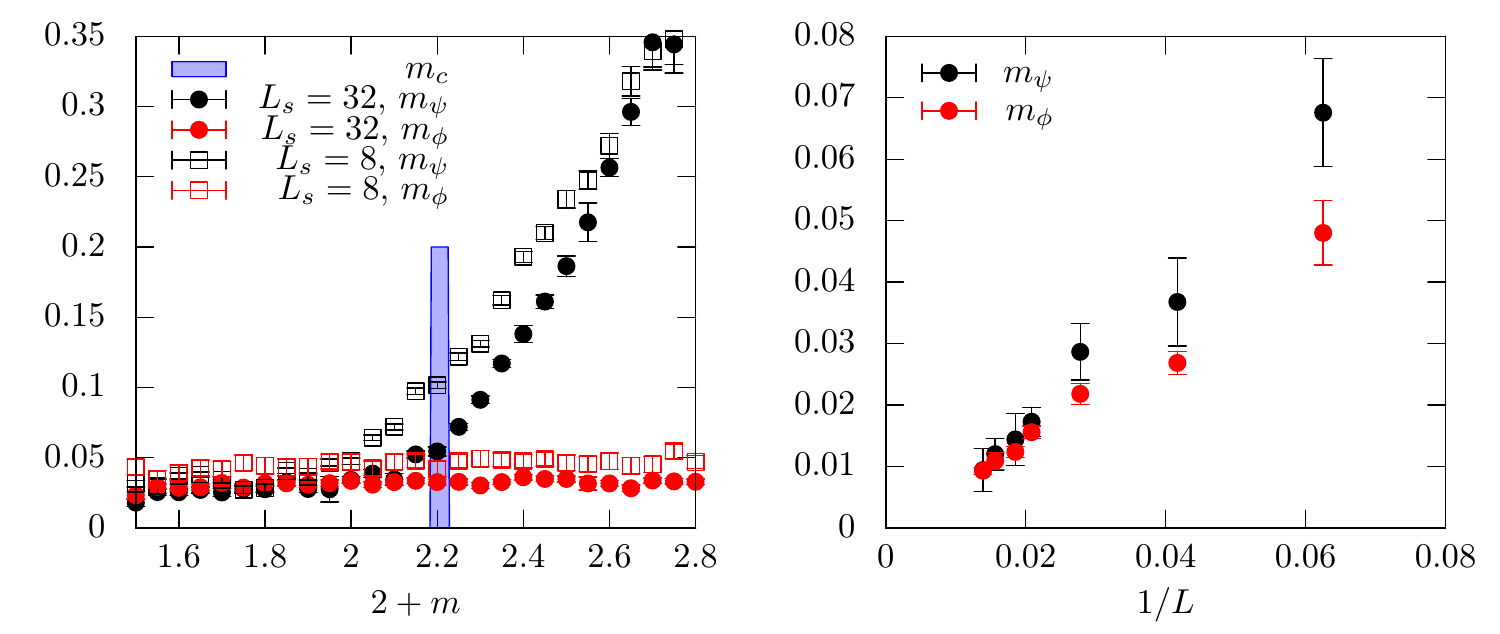}%
\caption{{\it Left plot}: Boson and fermion masses at coupling $g=0.5$
  as a function of the bare mass for two different lattice
  volumes. {\it Right plot}: Boson and fermion masses for $g=0.5$ at the critical
  mass $m_c$ versus the inverse lattice extent.}
\label{fig_MassesVsbaremassSUSYON}
\end{center}
\vspace{-0.5cm}
\end{figure}

 Next we investigate the boson and fermion mass spectrum of the
 theory.  The fermion correlator can be measured with high accuracy
 using the open fermionic string update introduced in
 \cite{Wenger:2008tq}.  In the left panel of
 fig.~\ref{fig_MassesVsbaremassSUSYON} we show the observed masses at
 fixed coupling $g=0.5$ for the two volumes with spatial extent
 $L_s=8$ and 32 and temporal extent $L_t=32$.  The boson mass $m_\phi$
 depends noticeably on the volume, but is essentially independent of
 the bare fermion mass and remains constant across the phase
 transition. The fermion mass $m_\psi$ on the other hand drops when
 the bare mass is decreased towards its critical value and becomes
 degenerate with the boson mass in the regime close to the critical
 point $m_c$ (denoted by the blue bar).  The mass degeneracy is
 further examined in the right panel of
 fig.~\ref{fig_MassesVsbaremassSUSYON} where the two masses $m_\phi$
 and $m_\psi$ at the critical mass $m_c$ for $g=0.5$ are plotted
 versus the inverse lattice extent. The data indicate that the mass
 degeneracy between the boson and fermion mass at the critical point
 survives the continuum limit where the model becomes chirally
 invariant.

\begin{figure}[t]
\begin{center}
\includegraphics{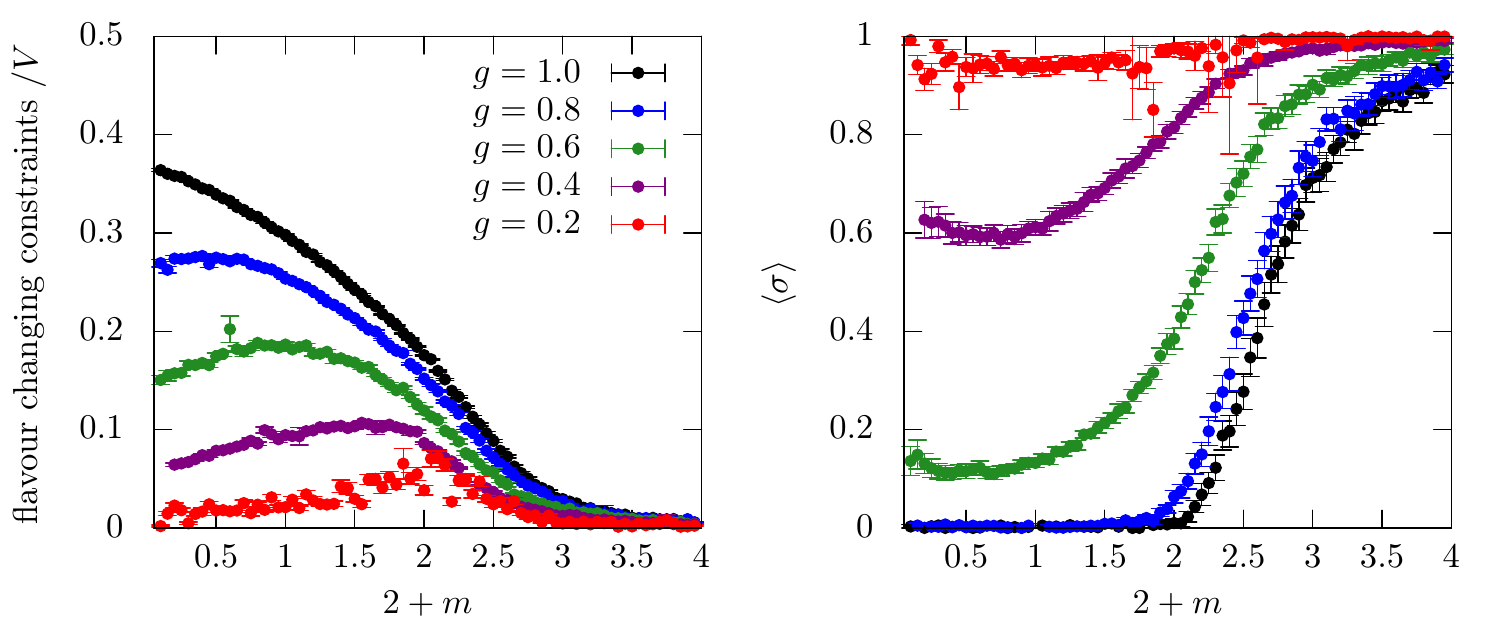}%
\caption{{\it Left plot}: Density of flavour changing constraints as a
  function of the bare mass for a range of couplings. {\it Right
    plot}: Same but for the average sign $\langle \sigma \rangle$.}
\label{fig_flavourchangincconstraintsAndZratios}
\end{center}
\vspace{-0.5cm}
\end{figure}

We emphasise again that the above calculations are made in the phase
quenched model. The only sources of negative contributions are the
flavour changing constraints
$\phi_r\phi_g\overline{\psi}_{r}\psi_{g}$. We are investigating their
occurrence in the left panel of
fig.~\ref{fig_flavourchangincconstraintsAndZratios} where the flavour
changing constraint density is plotted versus the bare mass for
various couplings $g$ on a lattice of size $8\times 8$. For large $g$
the number of flavour changing interactions grows for decreasing $m <
m_c$ while for smaller couplings the number remains rather small. On
the other hand, for large values of $m > m_c$ the density of flavour
changing interactions vanishes for any coupling. In this region there
are predominantly flavour diagonal constraints contributing positively
with $\phi^2_r$ or $\phi^2_g$. As a consequence the average sign
$\langle \sigma \rangle$ plotted in the right panel of
fig.~\ref{fig_flavourchangincconstraintsAndZratios} appears to behave
reasonably well and this suggests that approaching the chiral limit
from $m > m_c$ is under good control.

\section{Summary and outlook}
We constructed a fermion loop formulation of the supersymmetric
nonlinear O$(N)$ sigma model on an Euclidean space-time lattice using
the Wilson derivative for both the bosonic and fermionic fields. This
formulation maintains both the O($N$) symmetry and the constraints
arising from the supersymmetry and in principle allows a
straightforward simulation of the model. For $N=2$ we determined the
critical mass, defining the massless model in the continuum limit, for
a range of couplings and showed how the fermion and boson masses
become degenerate at this point. A careful analysis of Ward identities
is necessary in order to clarify whether or not the supersymmetry is
fully restored in the continuum limit, but in principle it is now
possible to determine the full phase diagram of the model, i.e., the
range of parameters for which the supersymmetry is spontaneously
broken, if at all.

One obstacle towards this goal is the fluctuating sign stemming from
the bosonic degrees of freedom entering in the constraints
$\phi\psi=0$. However, it is not clear whether this sign problem
survives the continuum limit since it is a pure artefact of the
particular regularisation chosen.  In fact, the sign problem can be
completely avoided by introducing the Wilson term only in the
fermionic sector and performing the hopping expansion also for the
bosonic degrees of freedom as outlined in section
\ref{sec:loop_formulation}, but of course it remains to be seen
whether supersymmetry is restored in the continuum also for such a
regularisation.

\bibliographystyle{JHEP}
\bibliography{susynonlinONsigmaproceedings}

\end{document}